\documentclass[preprint,1p,11pt]{ISAS_IR}

\usepackage[english]{babel}
\usepackage[latin1]{inputenc}
\usepackage{calc}
\usepackage{amsmath}
\usepackage{amssymb}
\usepackage{microtype}
\usepackage{lmodern}

\input{ISASMatheMacros.sty} 
\input{Defs.h}

\usepackage{tikz}
\usepackage{pgfplots}
\usepackage{booktabs}
\usepackage{subfigure}

\usetikzlibrary{positioning}
\usetikzlibrary{shapes,shadows}

\tikzstyle{block} = [draw, fill=blue!10, rectangle, 
    minimum height=2em, minimum width=2em, text width= 4.5em, text centered, drop shadow, rounded corners]
\tikzstyle{entry} = [draw, rectangle, 
    minimum height=2em, minimum width=2em, text width=3em, text centered]
\tikzstyle{pinstyle} = [pin edge={to-,thin,black}]

\begin{document}

\begin{frontmatter}

\title{Sequence-Based Control for \\ Networked Control Systems  \\ Based on Virtual Control Inputs}

\author[isas]{Achim Hekler}
\ead{Achim.Hekler@kit.edu}

\author[isas]{J\"org Fischer}
\ead{J\"org.Fischer@kit.edu}

\author[isas]{Uwe~D.~Hanebeck}
\ead{Uwe.Hanebeck@ieee.org}

\address[isas]{Intelligent Sensor-Actuator-Systems Laboratory (ISAS)\\
Institute for Anthropomatics\\
Karlsruhe Institute of Technology (KIT), Germany}

\begin{abstract}
In this paper, we address the problem of controlling a system over an unreliable connection that is affected by time-varying delays and randomly occurring packet losses.
A novel sequence-based approach is proposed that extends a given controller designed without consideration of the network-induced disturbances.
Its key idea is to model the unknown future control inputs by random variables, the so-called \emph{virtual control inputs}, which are characterized by discrete probability density functions.
 Subject to this probabilistic description, the actual sequence of future control inputs is determined and transmitted to the actuator.
 The high performance of the proposed approach is demonstrated by means of Monte Carlo simulation runs with an inverted pendulum on a cart 
 and by a detailed comparison to standard NCS approaches.

\end{abstract}

\end{frontmatter}

%
%

\section{Introduction}
\noindent
In networked control systems (NCS), communication between components of the control loop can be realized via a communication network instead of a transparent connection \cite{Yang06,Hespanha07}. 
This system architecture offers many advantages, such as simple installation and maintenance, as well as a high flexibility in the system structure. 
Therefore, NCS can already be found in a wide range of applications, e.g., unmanned vehicles \cite{Seiler05}, telepresence systems \cite{Hirche04}, or mobile sensor networks \cite{Ogren04}. 
 
However, it is well known that compared to a transparent connection, the presence of a communication network in the control loop decreases the quality of control or even destabilizes the system \cite{Zhang01, Bemporad10, Heemels10}. 
This is mainly caused by time-varying transmission delays and randomly occurring packet losses, limited bandwidth of the communication channel, or quantization errors. 
Consequently, control methods for NCS have to consider both communication and control aspects. 

\begin{figure}
   \centering
     \tikzstyle{block} = [draw, fill=blue!10, rectangle, 
    minimum height=2em, minimum width=2em, text width= 4.5em, text centered, drop shadow]
\tikzstyle{blockoS} = [draw, fill=blue!10, rectangle, 
    minimum height=1em, minimum width=2em, text width= 4.5em, text centered]
\tikzstyle{entry} = [draw, rectangle, 
    minimum height=2em, minimum width=2em, text width=3em, text centered]
\tikzstyle{pinstyle} = [pin edge={to-,thin,black}]

\begin{tikzpicture}[auto, node distance=2.5cm]
    \node [block] (controller) {Controller};
    \node [block, xshift = - 2.2cm, below = 2 cm of controller] (plant) {Plant};
    \node [block, right = 3cm of plant](actuator) {Actuator};
    \node [blockoS, above = 0cm of actuator] (buffer) {Buffer}; 
    \node[rectangle, draw, fill=red!20, aspect=3, minimum height = 2em, minimum width = 2em, above = 0.9cm of actuator, drop shadow] (network){Network};
    \draw [->,thick] (actuator) -- node [above,name=ud] {$\rvec{u}_k$}(plant);
    \draw [->,thick] (plant) |- node [near start,name=x] {$\underline{x}_k$}(controller);
    \draw[->,thick](controller) -| node [near end,name=u] {$U_k$}(network);
    \draw[->,thick](network) -- (buffer);
\end{tikzpicture}
     \caption{Considered NCS architecture. Controller and actuator are connected through a communication network, whereas the link between sensor and controller is transparent.
             For compensation of time delays and packet losses, a controller is employed, which transmits a whole sequence $U_k$ of optimal control inputs instead of just a single one.}
      \label{fig:NCSsystem}
\end{figure}
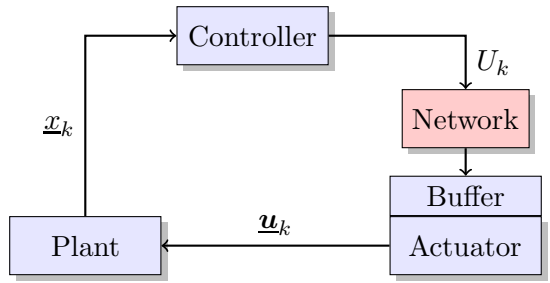

Our approach is based on the well-established control technique, named sequence-based control, which does not transmit just a single control input but a whole sequence of reasonable inputs for the future time steps.
This idea takes advantage of the property of modern communication networks, in which data is transmitted in large time-stamped packets.
The successfully transmitted sequences are stored in a buffer at the actuator and a specific selection logic enables that some reasonable inputs can be passed on to the plant at every time step.

The main problem of this procedure is that the controller actually has to know the control inputs applied by the actuator in the past and in the future, 
in order to be able to determine the current optimal sequence of control inputs. 
However, this demand is obviously not realizable, due to unpredictability of the network-induced delays. 

\subsection{Related Work} 
In \cite{Gruene09} and \cite{Polushin08}, a deterministic protocol is proposed to guarantee that the sequence of control inputs used by the controller for state predictions coincides with the sequence applied by the actuator. 
By enforcing this property, the so-called \emph{prediction consistency}, there are some significant drawbacks.
Especially in the case of long time delays, the controller is frequently in the so-called recovery mode, in which 
the actuator rejects inconsistently predicted sequences, even when they are based on recent measurements.

In \cite{Quevedo08}, a scenario-based NCS controller is proposed calculating the optimal control inputs for each possible delay of the previously transmitted sequences.
Then, the set of control sequences is transmitted to a smart actuator, which selects the correct sequence. 
Obviously, the complexity of this approach increases strongly when longer time delays occur.

Many sequence-based NCS controller neglect communication aspects and send at every time step a sequence of inputs resulting from an open-loop control problem exclusively depending on the current system state, e.g., see \cite{Quevedo11} or \cite{Quevedo07}.
Consequently, these approaches do not incorporate into the control decision the control inputs sent by the controller in the past and stored in the buffer of the actuator.
But, these inputs also have potentially an effect on the future evolution of the system.

A feedback controller designed for systems without network-induced disturbances is employed in \cite{Liu07} for 
determining a control input sequence.
Along with a deterministic model of the plant, future states and therefore, future control inputs are predicted, which are then transmitted in a control input sequence to the actuator.

\subsection{Key Idea}
In this paper, we extend any given state feedback controller designed without considering network-induced disturbances.
The novel idea is to model the future inputs by random variables, named \emph{virtual control inputs}.
These random variables are characterized by  discrete probability density functions over potential control inputs, which are derived from the data transmitted by the controller in the past.
Based on this probabilistic description representing the best knowledge of the controller about the situation at the actuator, a sequence of control inputs for the future time steps is determined.

\subsection{Notation}
Throughout the paper, random variables $\rv{a}$ are written in bold face letters, whereas deterministic quantities $a$ are in normal lettering.
Furthermore, the notation $\rv{a} \sim f(a)$ means that the random variable $\rv{a}$ is characterized by its probability density function $f(a)$.
A vector-valued quantity $\vec{a}$ is indicated by underlining the corresponding identifier and matrices are always referred to with bold face capital letters, e.g., $\mat{A}$.
The notation $a_k$ refers to the quantity $a$ at time step $k$. 
Furthermore, $a_{k \vert t}$ denotes the quantity $a$ at time step $k$ based on information up to time $t$. 
The term $\mat{0}_{m \times r}$ refers to a matrix with dimension $m \times r$ with all entries equal to 0 and $\mat{0}_r$ is an abbreviation for $\mat{0}_{r\times r}$. Finally, the term $\vec{0}_r$ means $\mat{0}_{r \times 1}$ and $\mat{I}_n$ denotes the identity matrix of dimension $n \times n$.

\subsection{Outline}
The remainder of the paper is organized as follows:
In the next Section, the problem is defined and the assumptions made are listed. 
Then, the proposed controller scheme for NCSs is described in detail and stability properties are examined.
Section~VI presents  simulation results with an inverted pendulum and compares the proposed approach to standard NCS techniques.
A summary and an outlook on future work concludes the paper.


\section{Considered Problem} \label{Sec_Section}
\noindent
Throughout the paper, we consider a discrete-time linear dynamic system described in state-space form via
\begin{equation}
  \rvec{x}_{k+1} = \mat{A} \rvec{x}_k + \mat{B} \rvec{u}_k + \rvec{w}_k \ ,
  \label{eq:system}
\end{equation}
where $\rvec{x}_{k} \in \mathbb{R}^s$ denotes the system state at time step $k$ and $\rvec{u}_k \in \mathbb{R}^n$ the control input applied by the actuator.
Note that due to time delays and packet losses in the network, $\rvec{u}_k$ is a random variable.
The system noise is subsumed by $\rvec{w}_k \sim f^w(\vec{w}_k)$ and is assumed to be a zero-mean Gaussian noise process. 
Furthermore, the system matrices $\mat{A} \in \mathbb{R}^{s \times s}$ and $\mat{B} \in \mathbb{R}^{s \times n}$ are assumed to be known.

The components of the control loop are time-triggered, synchronized, and have identical cycle times. 
Furthermore, we assume that the actuator does not have sufficient calculation capacity to perform local control.  

In this paper, we restrict our considerations to the case, where the communication network is solely present in the controller-to-actuator link.
We further assume that the controller has perfect information about the current system state $\vec{x}_k$ of the plant, i.e., the system state is completely measurable by the sensor
and the connection between sensor and controller is perfect.

The employed network is capable of transmitting large time-stamped data packets and does not provide acknowledgements for successfully transmitted data as in so-called UDP-like protocols.
The data transmission might be subject to time-varying and possibly unbounded\footnote{By allowing the time-delays to be unbounded, packet losses are incorporated into the description of the random delay processes since the loss of a packet corresponds to an infinite time-delay}
delays, modeled as a discrete random process $\rv{\tau}_k \in \mathbb{N}$. The realizations of this process describe how many time steps a packet generated in time step $k$ will be delayed until it is received.
It is assumed that $\rv{\tau}_k \sim f^\tau({\tau})$ is a white stationary process that is independent of $\rv{w}_k$ and that the probability density function $f^\tau({\tau})$ is known.

Finally, we assume that a controller with a linear state feedback control law
\begin{equation}
 \vec{u}_k = \mat{L} \cdot \vec{x}_k
 \label{eq:givenController}
\end{equation}
is given that is designed without consideration of the network-induced disturbances. In the following, we propose a scheme that extends this given controller in such a way that it can deal with time delays and packet losses.


\section{Sequence-based Control}\label{Sec:Packet}
\noindent
In this section, we briefly review the general concept of sequence-based control as, e.g., used in \cite{Bemporad98,Gruene09,Quevedo11, Liu06, Tang07, Liu07}, 
since our control approach presented in the next section is based on this fundamental control concept.

In sequence-based control, a controller generates not just a single control input for the current control cycle, but also control inputs for future $N$ time steps (with $N \in\mathbb{N}$). 
The whole control input sequence is lumped into one data packet and sent over the network to the actuator. 
The actuator is equipped with a buffer, in which the most recent control input sequence is stored, i.e., that sequence that has the highest time stamp among all received packets.
Therefore, when a new packet is received by the actuator, it is taken into the buffer if its time stamp is higher than the one of the packet already stored in the buffer, otherwise it is neglected.
Finally, in every time step, the actuator applies the appropriate control input of the buffered sequence to the plant, i.e., that control input of the sequence that corresponds to the current time step.

For the following derivations, we need some further notations. 
A control input sequence generated by the controller at time $k$ will be denoted by $U_k$. 
An entry of that packet is denoted by $\vec{u}_{k+m|k}$ with $m \in \{0, 1, ..., N\}$, where the first part of the index (here: $k+m$) gives the time step, for which the control input is intended to be applied to the plant. 
The second part of the index (here: $k$) specifies the time step, when the control input was generated. 
For a packet of length $N + 1$ generated in time step $k$, this gives
\begin{equation}
	U_k = \{\vec{u}_{k|k}, \vec{u}_{k+1|k}, \dots, \vec{u}_{k+N|k}\} \ . 
\end{equation}

For example, let us assume the controller packet $U_k$ is received by the actuator at time step $k+\tau_k$ with $\tau_k \in \mathbb{N}$. 
If none of the packets 
\begin{equation}
	U_{k+1}, U_{k+2}, \dots, U_{k+\tau_k}
  \label{eq:arrivedPackets}
\end{equation}
has been received by the actuator so far, then the buffer is overwritten with the entries of $U_k$ and the input $\vec{u}_{k+\tau_k \vert k}$ is applied to the plant.
Otherwise, if the actuator has received any packet from (\ref{eq:arrivedPackets}) until time step $k+\tau_k$, say, e.g.,  $U_{k+i}$, for $i=1,\dots \tau_k$, $U_k$ is neglected and $\vec{u}_{k+\tau_k \vert k+i}$ of the buffered sequence $U_{k+i}$ is applied.

Since we do not assume that the time delays are bounded, it may happen that the buffer runs empty. 
In this case, the controller operates with a default input $\vec{u}^d$. 

It is obvious that the control inputs applied by the actuator depend on the packet delays as well as losses and, therefore, inherit the stochastic nature of the network. This gives rise to the stochastic control approach discussed in the next section.

\section{Sequence-based Control with Virtual Control Inputs}\label{Sec:Controller}
\noindent
It should be obvious from the preceding section that in sequence-based control, control inputs from packets sent in previous time steps may actively affect the future evolution of the plant. 
It therefore seems only reasonable to take these old control inputs explicitly into account, when calculating new control inputs, which is also the main idea of the proposed approach. 

In the following, we derive in Sec. \ref{sub:VCI} a stochastic description of these already transmitted, but possibly applied control inputs, that we call virtual control inputs. Then, utilizing the concept of virtual control inputs, we design the controller in Sec. \ref{sub:ContrDesign}.

\subsection{Virtual Control Inputs}
\label{sub:VCI}
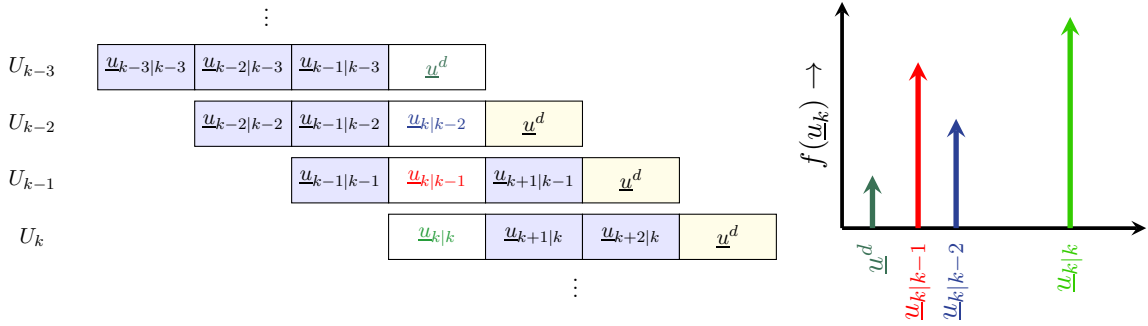
\begin{figure}
 \begin{minipage}{0.6\textwidth}
   \tikzstyle{entry} = [draw, rectangle, fill=blue!10,
    minimum height=0.8cm, minimum width=1.7cm, text centered]
\tikzstyle{sentry} = [draw, rectangle, 
    minimum height=0.8cm, minimum width=1.7cm, text centered]
\tikzstyle{pinstyle} = [pin edge={to-,thin,black}]
\tikzstyle{dentry} = [draw, rectangle, fill=yellow!10,
    minimum height=0.8cm, minimum width=1.7cm, text centered]

\begin{tikzpicture}[scale=0.75, transform shape]
\node [entry] at (-6.5,4.5) {$\underline{u}_{k-3 \vert k-3}$};
 \node [entry] at (-4.8,4.5) {$\underline{u}_{k-2 \vert k-3}$};
 \node [entry] at (-3.1,4.5) {$\underline{u}_{k-1 \vert k-3}$};
 \node [sentry] at (-1.4,4.5) {\color{rgb:red,2;green,4;blue,3}$\underline{u}^d$};
 \node [entry] at (-4.8,3.5) {$\underline{u}_{k-2 \vert k-2}$};
 \node [entry] at (-3.1,3.5) {$\underline{u}_{k-1 \vert k-2}$};
\node [sentry] at (-1.4,3.5) {\color{rgb:red,2;green,3;blue,7}$\underline{u}_{k \vert k-2}$};
\node [dentry] at (0.3,3.5) {$\underline{u}^d$};
 \node [entry] at (-3.1,2.5) {$\underline{u}_{k-1 \vert k-1}$};
  \node [sentry] at (-1.4,2.5) {\color{rgb:red,2;green,0;blue,0} $\underline{u}_{k \vert k-1}$};
  \node [entry] at (0.3,2.5) {$\underline{u}_{k+1 \vert k-1}$};
  \node [dentry] at (2,2.5) {$\underline{u}^d$};
  \node [sentry] at (-1.4,1.5) {\color{rgb:red,2;green,8;blue,3}$\underline{u}_{k \vert k}$};

  \node [entry] at (0.3,1.5) {$\underline{u}_{k+1 \vert k}$};
  \node [entry] at (2,1.5) {$\underline{u}_{k+2 \vert k}$};
  \node [dentry] at (3.7,1.5) {$\underline{u}^d$};
 \node  at (-8.5,4.5) {$U_{k-3}$};
 \node  at (-8.5,3.5) {$U_{k-2}$};
 \node  at (-8.5,2.5) {$U_{k-1}$};
 \node  at (-8.5,1.5) {$U_{k}$};
 \node at (-4.3873,5.4856) {$\vdots$};
 \node at (1.0257,0.7053) {$\vdots$};
\end{tikzpicture}
  \end{minipage}
  \hspace{0.3cm}
 \begin{minipage}{0.4\textwidth}
   \begin{tikzpicture}[>=stealth]
  \draw[->, line width = 1.4pt] (1,0) -- (1,3);
  \draw[->, line width = 1.4pt] (1,0) -- (5,0);
  \draw[->, color={rgb:red,2;green,0;blue,0}, line width = 2pt] (2.0,0) -- (2.0,2.2);
  \draw[->, color={rgb:red,2;green,8;blue,0},  line width = 2pt] (4.0,0) -- (4.0,2.8);
  \draw[->, color={rgb:red,2;green,3;blue,7},  line width = 2pt] (2.5,0) -- (2.5,1.45);
  \draw[->, color={rgb:red,2;green,4;blue,3},  line width = 2pt] (1.4,0) -- (1.4,0.7);
  \node[rotate=90] at (0.63, 1.5) {$f(\underline{u}_{k}) \rightarrow$};
  \node[rotate=90] at (2.0,-0.7) {\color{rgb:red,2;green,0;blue,0} $\underline{u}_{k \vert k-1}$};
  \node[rotate=90] at (4.0,-0.52) {\color{rgb:red,2;green,8;blue,0} $\underline{u}_{k \vert k}$};
  \node[rotate=90] at (2.5,-0.7) {\color{rgb:red,2;green,3;blue,7} $\underline{u}_{k \vert k-2}$};
  \node[rotate=90] at (1.4,-0.4) {\color{rgb:red,2;green,4;blue,3} $\underline{u}^d$};
\end{tikzpicture}
 \end{minipage}
   \caption{Schematic illustration of the transmitted packets. Control inputs corresponding to the same time step are vertically aligned. 
   For example, the potential control inputs that could be applied by the actuator at time step $k$ are indicated by white rectangles. 
   The last yellow entry is the default value that would be employed if the buffer runs empty.}
   \label{fig:packets}
\end{figure}

In this section, we introduce the novel concept of virtual control inputs. To that end, we first define the information set $\mathcal{I}_k$ that summarizes the information the controller can use at time step $k$ to calculate $U_k$. Considering causal controllers, the information set includes all measurements 
and all control packets that were received and sent,  respectively, until time step $k$.
Furthermore, the information set contains the information about the given feedback matrix $\mat{L}$, the dynamics of the system $\mathcal{D}$ given by Eq. (\ref{eq:system}), the buffering logic $\mathcal{B}$ of the actuator described in Sec. \ref{Sec:Packet}, and the stochastic characteristics of the process and measurement noise, so that
\begin{equation}
	\mathcal{I}_k = \{\vec{x}_0, \ \hdots \ , \vec{x}_k, U_{0}, \ \hdots \ , U_{k-1}; \mat{L}, \mathcal{D}, \mathcal{B}, f^w(\vec{w}), f^\tau(\tau) \} \ .
\end{equation} 
\begin{Remark}
	The information that can be used by the controller, i.e., $\mathcal{I}_k$, does not contain the knowledge that the controller will sent control input sequences in future. This means that the controller does not consider that in the next time step the state will be measured and a control sequence will be generated and sent to the actuator. This is justified by the fact that the proposed controller must not use this information for two reasons. First, the controller has to generate open-loop control sequences since the actuator has no access to state measurements. Second, if an entry of the control input sequence $U_{k+1}$ is applied by the actuator, it is not possible that a control input of the sequence $U_{k}$ will be applied afterwards. Therefore, the influence of future control sequences $U_{k+1}$ (and following sequences) must not be considered in the generation of the control sequence $U_k$.
\end{Remark}
Based on $\mathcal{I}_k$, we define the virtual control inputs as follows.
\begin{Definition}[Virtual Control Inputs]
A virtual control input $\rvec{u}_{k+m|k}^v \sim f(\vec{u}_{k+m|k}^v)$ is a random variable that characterizes the control input $\rvec{u}_{k+m}$ based on the information $\mathcal{I}_k$ (with $k, m \in \mathbb{N}$).
\end{Definition}
\begin{Remark}
	It is important to distinguish 1) $\vec{u}_{k+m}$, that is a realization of the control input $\rvec{u}_{k+m}$, 2) $\vec{u}_{k+m|k}$, that is an entry of the packet $U_k$ and that describes the control input applicable at time step $k+m$ calculated by the controller at time step $k$, 3) the virtual control input $\rvec{u}_{k+m|k}^v$, that is a prediction of $\rvec{u}_{k+m}$ based on $\mathcal{I}_k$, and 4) $\vec{u}_{k+m|k}^v$ that is realization of $\rvec{u}_{k+m}$.
\end{Remark}
To derive the probability density function $f(\vec{u}_{k+m|k}^v)$ of the virtual control inputs $\rvec{u}_{k+m}^v$, we note that, based on the information set $\mathcal{I}_k$, there is only a finite set of discrete values of control inputs that could be applied by the actuator. This is illustrated in Fig.~\ref{fig:packets} for the case of $N = 2$, where the control inputs possibly applied at time step $k$ are marked by white rectangles. It should be noted that, although this finite set of control inputs is discrete, the control inputs itself are over a continuous domain.
The structure of the uncertainty can formally be described by a Dirac mixture density, so that it holds for the probability density functions of the virtual control inputs
\begin{align*}
  f(\vec{u}_{k+m|k}^v) & = \rv{\alpha}^{(N-m+1)}_{k+m|k} \ \delta(\vec{u}^v_{k+m|k} - \vec{u}^d) +  
  \displaystyle\sum\limits_{i=0}^{N-m}\rv{\alpha}^{(i)}_{k+m|k}\;\delta(\vec{u}^v_{k+m|k} - \vec{u}_{k+m|k-i} ) \ ,
\end{align*}
with
\begin{eqnarray*}
	m \in \{0, 1, \cdots, N\} \ , \ \ \ \ \	\displaystyle\sum\limits_{i=0}^{N-m+1} \rv{\alpha}^{(i)}_{k+m|k} = 1 \ , 
\end{eqnarray*}
where $\delta()$ is the Dirac delta function and $\rv{\alpha}^{(i)}_{k+m|k}$ are scalar weighting factors. 
\begin{Remark}
	Since we do not make the assumption that the delays are bounded, it can occur that the actuator runs out of applicable control inputs. 
	This is taken into account in $f(\vec{u}_{k+m|k}^v)$ by the term $\vec{u}^d$.
\end{Remark}
The weighting factors express the probability that the corresponding control input $\vec{u}_{k+m|k-i}$ is applied by the actuator, i.e.,
\begin{equation*}
  \rv{\alpha}^{(i)}_{k+m|k} = {\rm Prob}(\rvec{u}_{k+m} = \vec{u}_{k+m|k-i} | \mathcal{I}_k) \ .
\end{equation*}
The control input $\vec{u}_{k+m|k-i}$ is applied by the actuator if the sequence buffered in the actuator at time step $k+m$ has been generated by the controller $k+m-(k-i)=m+i$ time steps ago. 
In other words, $\vec{u}_{k+m|k-i}$ is applied by the actuator if the \textit{age} of the buffered sequence, i.e., the difference between time step of generation and actual time step, at time step $k+m$ is equal to $m + i$. In the following, we denote the age of the buffered sequence at time step $k$ by $\rv{\theta}_k$. With this notation it holds that
\begin{equation*}
  \rv{\alpha}^{(i)}_{k+m|k} = {\rm Prob}(\rv{\theta}_{k+m} = i| \mathcal{I}_k) \ .
\end{equation*}
Therefore, the weighting factors can be interpreted as estimates of $\rv{\theta_k}$. It is shown in the appendix that $\rv{\theta}_k$ can be formulated as the state of a Markov chain (with state space $\{0, 1, 2, ..., N + 1\}$) that is governed by the transition matrix $\mat{P}$ for that holds
%
\begin{align}
	\mat{P} = & \label{eq:P}
	\begin{bmatrix}
		p_{(0,\ 0)} & p_{(0,\ 1)} & 0      & 0  & \cdots & 0\\
		p_{(1,\ 0)} & p_{(1,\ 1)} & p_{(1,\ 2)} & 0  & \cdots & 0\\
		p_{(2,\ 0)} & p_{(2,\ 1)} & p_{(2,\ 2)} & p_{(2,\ 3)} & \cdots & 0\\
		\vdots & \vdots & \vdots & \vdots & \ddots & \vdots\\
		\vdots & \vdots & \vdots & \vdots & \vdots & p_{(N,\ N+1)}\\
		p_{(N+1,\ 0)} & \ p_{(N+1,\ 1)} & \ p_{(N+1,\ 2)} & \ p_{(N+1,\ 3)} & \ \cdots & \ p_{(N+1,\ N+1)}
	\end{bmatrix}\ , 
\end{align}
where the $p_{(i,\ j)}$ are equal to ${\rm Prob}\left(\rv{\theta}_{k+1} = j|\rv{\theta}_k = i \right)$ and can be calculated by
\begin{eqnarray*}
	p_{(i, \ j)} = 
	\begin{cases}	
		0 & \text{for} \ j \geq i + 2 \ , \\
		1 - \sum\limits_{r=0}^{i} q_r & \text{for} \ j = i + 1 \ , \\ 
    q_j& \text{for} \ j \leq i \ .
  \end{cases}
\end{eqnarray*}
%
Thereby, $q_i$ describes the probability of the event that a packet is delayed by $i$ time steps. These probabilities can easily be derived since the probability density function of the time delays of the network connection are known.
Arranging the weighting factors $\rv{\alpha}_{k+m|k}^{(i)}$ in form of a vector
\begin{equation}
	\rvec{\alpha}_{k+m|k} = \left[ \rv{\alpha}^{(0)}_{k+m|k}, \cdots, \rv{\alpha}^{(N-m+1)}_{k+m|k} \right]^{\rm T} \ ,
\end{equation}
it holds that 
\begin{equation}
	\rvec{\alpha}_{k+m|k} = \frac
	{
		\begin{bmatrix} 
			\mat{I}_{N+2-m}  & \mat{0}_{(N+2-m)\times m}
		\end{bmatrix}
		\left(\mat{P}^m\right)^{\rm T} \cdot \rvec{\alpha}_{k|k} 
	} 
	{ 
		{\rm Tr}\left( {\rm diag} \left[
			\begin{bmatrix} 
				\mat{I}_{N+2-m}  & \mat{0}_{(N+2-m)\times m}
			\end{bmatrix} \left(\mat{P}^m\right)^{\rm T} \cdot \rvec{\alpha}_{k|k} \right]
		\right)
	} 
	\label{eq:alphaPred}\ ,
\end{equation}
where ${\rm diag} \left[\vec{s}\right]$ denotes a matrix with the elements of the vector $\vec{s}$ on the diagonal and zeros everywhere else, and the term ${\rm Tr}\left(\mat{S}\right)$ denotes the trace of $\mat{S}$. The term in the numerator represents a $m$-step future prediction of $\theta_k$, where only the first $N+2-m$ elements are kept. The other elements describe 
the probability that future control inputs are applied, which are based on information that is not available, e.g., $\mathcal{I}_{k+1}$ and is therefore not used. The denominator normalizes the extracted subset of the predicted vector so that it sums up to one.

According to Eq. (\ref{eq:alphaPred}), if $\rvec{\alpha}_{k|k}$ is known, the other weighting factors $\rvec{\alpha}_{k+1|k} \cdots \rvec{\alpha}_{k+N|k}$ can be derived by means of the transition matrix $\mat{P}$. To derive $\rvec{\alpha}_{k|k}$ we note that the state $x_k$ can also be interpreted as the continuous-valued outputs of a \textit{Hidden Markov Model}, that is governed by $\rv{\theta}_k$. Hence, it is possible to apply the continuous-valued version of the Wonham filter \cite{Wonham64} in the form
%
\begin{equation}
	\rvec{\alpha}_{k|k} = \frac{\mat{H} \cdot \rvec{\alpha}_{k|k-1}}{{\rm Tr}\left[\mat{H}\right]}\ ,
\end{equation}
\begin{equation}
	\mat{H} = {\rm diag} \left[
		\begin{array}{c}
			f^w\left(\vec{x}_k - \mat{A}\vec{x}_{k-1} - \mat{B}\vec{u}_{k|k}\right)\\ 
			f^w\left(\vec{x}_k - \mat{A}\vec{x}_{k-1} - \mat{B}\vec{u}_{k|k-1}\right)\\
			\vdots \\
			f^w\left(\vec{x}_k - \mat{A}\vec{x}_{k-1} - \mat{B}\vec{u}_{k|k-N}\right)\\
			f^w\left(\vec{x}_k - \mat{A}\vec{x}_{k-1} - \mat{B}\vec{u}^d\right)
		\end{array} \right] \ .
\end{equation}
The prediction $\rvec{\alpha}_{k|k-1}$ can be calculated using Eq. (\ref{eq:alphaPred}).
It can be seen that $\rvec{\alpha}_{k|k}$ (and therefore $\rvec{\alpha}_{k+m|k}$) is time-varying. To reduce the complexity in the calculation, the weighting factors $\rvec{\alpha}_{k|k}$ can be approximated by its stationary probability solution ${\rm \lim_{k \rightarrow \infty}}\rvec{\alpha}_{k|0} = \vec{\alpha}_\infty$. The stationary solution $\vec{\alpha}_\infty$ can be computed by the equilibrium equation
\begin{equation}
	\vec{\alpha}_\infty = \mat{P}^{\rm T} \vec{\alpha}_\infty \label{eq:alphaInfty} 
\end{equation}
that always has a unique solution according to Markov chain theory. Using $\vec{\alpha}_{\infty}$ instead of $\rvec{\alpha}_{k|k}$ has the advantage that all weighting factors become time-invariant and the controller is easier to calculate. 
Furthermore, the stability of the closed-loop system can be analyzed more easily (see Sec. \ref{Sec:Stability}). 
We, therefore, derive the controller for both cases, but concentrate in the stability analysis on the time-invariant approximation.

In section \ref{sub:ContrDesign}, we will need the expected value of the virtual control inputs. These can be calculated by
\begin {align}
	E & \left\{  \rvec{u}_{k+m|k}^v \right\} = \displaystyle\int^\infty_{-\infty} \vec{u}_{k+m|k}^v \ f(\vec{u}_{k+m|k}^v) \: \mathrm d \vec{u}_{k+m|k}^v\nonumber\\
	& = \displaystyle\int^\infty_{-\infty} \vec{u}_{k+m|k}^v \left(
		\rv{\alpha}^{(N-m+1)}_{k+m|k} \ \delta(\vec{u}^v_{k+m|k} - \vec{u}^d) +  
  \displaystyle\sum\limits_{i=0}^{N-m}\rv{\alpha}^{(i)}_{k+m|k}\;\delta(\vec{u}^v_{k+m|k} - \vec{u}_{k+m|k-i} )\right) 
		\mathrm d\vec{u}_{k+m|k}^v \nonumber\\
	& = \rv{\alpha}^{(N-m+1)}_{k+m|k} \ \vec{u}^d + \displaystyle\sum\limits_{i=0}^{N-m} \rv{\alpha}^{(i)}_{k+m|k}\; \vec{u}_{k+m|k-i} \label{eq:expOfu} \  \ .
\end{align}
With the steady state approximation this becomes
\begin{equation}
 	E \left\{  \rvec{u}_{k+m|k}^v \right\} \approx  {\alpha}^{(N-m+1)}_\infty \ \vec{u}^d + \displaystyle\sum\limits_{i=0}^{N-m} {\alpha}^{(i)}_\infty\; \vec{u}_{k+m|k-i} \ \ .
\end{equation}

\subsection{Controller Design}
\label{sub:ContrDesign}
This subsection describes, how to design the sequence-based controller based on the linear feedback controller
\begin{equation}
	\hat{\vec{u}}_k = \mat{L} \cdot \vec{x}_k \ ,
\end{equation}
where the feedback matrix $\mat{L}$ was designed for the plant (\ref{eq:system}) without consideration of network effects by, e.g., pole placement or another modern control method, such as LQR, $\rm H_2$, or $\rm H_{\infty}$. 
In the following, we use the feedback matrix $\mat{L}$ to generate control input sequences based on the predicted future states of the plant.

Based on the measured state $\vec{x}_k$ at time step $k$, the entries of the control input sequence $U_k$ are calculated by
%
\begin{eqnarray}
	\vec{u}_{k|k} &=& \mat{L} \cdot \vec{x}_k\label{eq:Lx1} \ ,\\ 
	\vec{u}_{k+1|k} &=& \mat{L} \cdot E\{\rvec{x}_{k+1|k}\}\label{eq:Lx2} \ ,\\
  &\vdots&\nonumber\\
	\vec{u}_{k+N|k} &=& \mat{L} \cdot E\{\rvec{x}_{k+N|k}\label{eq:Lxm}\} \ ,
\end{eqnarray}
where $\rvec{x}_{k+m|k} \sim f(x_{k+m|\mathcal{I}_k})$ describes the state predictions conditioned on the information $\mathcal{I}_k$. The state predictions are random with respect to the process noise and the virtual control inputs. The expected value of the predicted state predictions $E\{\rvec{x}_{k+m|k}\}$ can be calculated by
\begin{align}
  E\{\rvec{x}_{k+1|k}\} &= E\{ \mat{A}\vec{x}_{k} + \mat{B} \rvec{u}_k + \rvec{w}_{k} \, |\mathcal{I}_k \} \nonumber \\
  	&= \mat{A}\vec{x}_{k} + \mat{B} \cdot E \{ \rvec{u}_{k} \, | \mathcal{I}_k \} + E \{\rvec{w}_k \, | \mathcal{I}_k\} \nonumber \\
	  &= \mat{A}\vec{x}_k +  \mat{B} \cdot E \{\rvec{u}_{k|k}^v \} \nonumber  \\
  	&= \mat{A}\vec{x}_{k} + \mat{B} \cdot \left(\rv{\alpha}^{(N+1)}_{k|k} \ \vec{u}^d + \displaystyle\sum\limits_{i=0}^{N} \rv{\alpha}^{(i)}_{k|k}\ \; \vec{u}_{k|k-i} \right) \nonumber \\
	  &\approx \mat{A}\vec{x}_{k} + \mat{B} \cdot \left({\alpha}^{(N+1)}_{\infty} \ \vec{u}^d + \displaystyle\sum\limits_{i=0}^{N} {\alpha}^{(i)}_\infty \; \vec{u}_{k|k-i} \right) \ , \nonumber \\\nonumber \\
  E\{\rvec{x}_{k+2|k}\} &= {E}\{ \mat{A} \rvec{x}_{k+1} + \mat{B}\rvec{u}_{k+1}  + \rvec{w}_{k+1} \, | \mathcal{I}_k \} \nonumber \\
  	&= \mat{A} \cdot {E}\{ \rvec{x}_{k+1} \, | \mathcal{I}_k \} + \mat{B} \cdot E\{\rvec{u}_{k+1} \, | \mathcal{I}_{k} \} + E \{\rvec{w}_{k+1} \, | \mathcal{I}_k\} \nonumber \\
  	&= \mat{A}\cdot E\{\rvec{x}_{k+1|k}\} +  \mat{B} \cdot E \{\rvec{u}_{k+1|k}^v \}\nonumber\\
  	&= \mat{A}\cdot E\{\rvec{x}_{k+1|k}\} + \mat{B} \cdot \left(\rv{\alpha}^{(N)}_{k+1|k} \ \vec{u}^d + \displaystyle\sum\limits_{i=0}^{N-1} \rv{\alpha}^{(i)}_{k+1|k}\ \; \vec{u}_{k+1|k-i} \right) \nonumber \\
		&\approx \mat{A} \cdot E\{\rvec{x}_{k+1|k}\} + \mat{B} \cdot \left({\alpha}^{(N)}_{\infty} \ \vec{u}^d + \displaystyle\sum\limits_{i=0}^{N-1} {\alpha}^{(i)}_\infty \; \vec{u}_{k+1|k-i} \right) \ , \nonumber \\\nonumber \\
	E\{\rvec{x}_{k+m|k}\}  &= E\{ \mat{A} \rvec{x}_{k+m-1} + \mat{B} \rvec{u}_{k+m-1} + \rvec{w}_{k+m-1} \, | \mathcal{I}_k \} \nonumber \\
	&= \mat{A} \cdot {E}\{ \rvec{x}_{k+m-1|k} \} + \mat{B} \cdot E\{\rvec{u}_{k+m-1|k}^v \}\nonumber \\
	&= \mat{A} \cdot {E}\{ \rvec{x}_{k+m-1|k} \} + \mat{B} \cdot \left(\rv{\alpha}^{(N-m+1)}_{k+m|k} \ \vec{u}^d + \displaystyle\sum\limits_{i=0}^{N-m} \rv{\alpha}^{(i)}_{k+m|k}\; \vec{u}_{k+m|k-i}\right)\nonumber  \\
	&\approx \mat{A} \cdot {E}\{ \rvec{x}_{k+m-1|k} \} + \mat{B} \cdot \left({\alpha}^{(N-m+1)}_\infty \ \vec{u}^d + \displaystyle\sum\limits_{i=0}^{N-m} {\alpha}^{(i)}_\infty \; \vec{u}_{k+m|k-i}\right) \ . \label{eq:pred}
\end{align}
\begin{Remark}
 If the size of the packets is equal to 1, i.e., $N=1$, then the proposed controller coincides with the given linear state feedback controller.
\end{Remark}
For taking the expected value, we use (\ref{eq:expOfu}) and the assumption that $\rvec{w}_k$ is zero-mean and independent of $\rvec{x}_k$ and $\rvec{u}_{k+m|k}^v$. 

From (\ref{eq:Lx1}) - (\ref{eq:Lxm}) and (\ref{eq:pred}) it follows that the extended controller using the time-invariant approximation of the weighting factors is linear not only in the measured state $\vec{x}_k$ but also in the control inputs of the sequences $U_k, \hdots, U_{k-N-1}$. 
Therefore, the controller can be formulated as linear feedback controller working on the augmented state
\begin{equation}
	U_k = \mat{\tilde{L}} \cdot \rvec{\psi}_k \ ,
 \label{eq:LSchlange}
\end{equation}
where the augmented state is defined by
\begin{equation}
	\rvec{\psi}_k = \left[ \vec{x}_k^{\rm T} \ \ \vec{\eta}_k^{\rm T} \right]^{\rm T} \ , 
	\label{eq:psi}
\end{equation}
with
\begin{equation}
	\begin{aligned}
		\vec{\eta}_k &=
		\begin{bmatrix}
			[\vec{u}_{k|k-1}^{\rm T} \ \ \vec{u}_{k+1|k-1}^{\rm T} \ \ \cdots \ \ \vec{u}_{k+N-1|k-1}^{\rm T} ] ^{\rm T}\\
			[\vec{u}_{k|k-2}^{\rm T} \ \ \vec{u}_{k+1|k-2}^{\rm T} \ \ \cdots \ \ \vec{u}_{k+N-2|k-2}^{\rm T} ]^{\rm T}\\
			\ \ \ \ \vdots\\
			\ \ \ \ \ \ [\vec{u}_{k|k-N+1}^{\rm T} \ \ \vec{u}_{k+1|k-N+1}^{\rm T}]^{\rm T}\\
			\ \ \ \ \ \vec{u}^{\rm T}_{k|k-N}
		\end{bmatrix} \in \mathbb{R}^{d} \ , \ \  \ \ \
		d = n \frac{N(N+1)}{2} \ .
	\end{aligned}
	\label{eq:eta}
\end{equation}

The vector $\vec{\eta}_k$ contains all control inputs of the already sent control input sequences $U_{k-1}$, $U_{k-2}$, $\hdots$, $U_{k-N}$ that still could be applied in time step $k$ or later.
The result (\ref{eq:LSchlange}) will be very useful in the next section, where a criterion for closed-loop stability for the extended controller is derived.

\section{Stability Issues}\label{Sec:Stability}
\noindent
In this section we will derive a criterion for closed-loop stability of the proposed controller. We concentrate on the case where the weighting factors of the virtual control inputs are approximated by its steady state distribution described by Eq. (\ref{eq:alphaInfty}). 
\begin{Remark}
It should be noted, that, although an approximation is used in the controller design, the derived stability results for this controller are exact and not approximated.
\end{Remark}
To derive the stability criterion, first, a model of network and actuator is derived, that, in a second step, will be combined with the model of the plant (\ref{eq:system}) and the controller (\ref{eq:LSchlange}).
%
\subsection{Combined Model of Network and Actuator}
Based on $\vec{\eta}_k$ and $\rv{\theta}_k$ as defined in section \ref{sub:VCI}, the combined state space model of network and actuator can be formulated as
\begin{eqnarray}
	\vec{\eta}_{k+1} & = & \mat{F} \vec{\eta}_k + \mat{G} U_k \label{eq:netwAndAct1} \ , \\
  \hat{\vec{u}}_k  & = & \mat{H}_{\theta_k} \vec{\eta}_k + \mat{J}_{\theta_k} U_k \label{eq:netwAndAct2} \ ,
\end{eqnarray}
with
\begin{align*}
	\mat{F} &= 
	\begin{bmatrix}
		\mat{0}_{nN\times n} & \mat{0}_{nN \times n(N-1)} & \mat{0}_{nN\times n} & \mat{0}_{nN\times n(N-2)} &\cdots & \mat{0}_{nN\times n} & \mat{0}_{nN\times n}\\
			\mat{0}_{n(N-1)\times n} & \mat{I}_{n(N-1)} & \mat{0}_{n(N-1)\times n} & \mat{0}_{n(N-1)\times n(N-2)} &\cdots & \mat{0}_{n(N-1)\times n} & \mat{0}_{n(N-1)\times n}\\
		\mat{0}_{n(N-2) \times n} & \mat{0}_{n(N-2) \times n(N-1)} & \mat{0}_{n(N-2)\times n} & \mat{I}_{n(N-2)} &\cdots & \mat{0}_{n(N-2) \times n} & \mat{0}_{n(N-2) \times n}\\
		\vdots & \vdots & \vdots & \vdots & \ddots & \vdots & \vdots \\
		\mat{0}_n & \mat{0}_{n \times n(N-1)} & \mat{0}_n & \mat{0}_{n \times n(N-2)} &\cdots & \mat{I}_n & \mat{0}_n \\
	\end{bmatrix} \ , \\\nonumber \  \\
	\mat{G} &= 
	\begin{bmatrix}
		\mat{0}_{nN \times n} & \mat{I}_{nN}\\
		\mat{0}_{n(N-1) \times n}  & \mat{0}_{n(N-1) \times nN}\\
		\vdots & \vdots\\
		\mat{0}_n & \mat{0}_{n \times nN}
	\end{bmatrix} \ , \ \ \ \ \ \ 
  \mat{J}_{\theta_k} =
	\begin{bmatrix}
	  \delta_{(\theta_k, 0)}\mat{I}_n & \mat{0}_{n \times nN}
	\end{bmatrix} \ , \ \ \ \ \ \ \ 
	\mat{H}_{\theta_k} =	
	\begin{bmatrix}
		\delta_{(\theta_k, 1)}\mat{I}_n \\ 
		\mat{0}_{n\times n(N-1)} \\  
		\delta_{(\theta_k, 2)}\mat{I}_n \\
		\mat{0}_{n\times n(N-2)} \\
		\vdots \\
		\delta_{(\theta_k, N)}\mat{I}_n  
		\end{bmatrix}^T ,
\end{align*}
where $\delta_{(\theta_k, i)}$ is the Kronecker delta function defined as
\begin{eqnarray*}
	\delta_{(\theta_k, i)} = 
	\left\{ \begin{array}{rcl}
		1 & \mbox{if} & \theta_k = i\\
		0 & \mbox{if} & \theta_k \neq i
	\end{array}\right. \ .
\end{eqnarray*}
%
\subsection{Model of the Closed-Loop System}
By using the augmented state $\rvec{\psi}_k$ form (\ref{eq:psi}) and combining (\ref{eq:system}), (\ref{eq:netwAndAct1}), and (\ref{eq:netwAndAct2}), it holds
\begin{align*}
\rvec{\psi}_{k+1} & = 
	\begin{bmatrix}
		\mat{A} & \mat{B} \cdot \mat{H}_{\theta_k}\\
		\mat{0}_{d\times s}  & \mat{F}
	\end{bmatrix}
	\rvec{\psi}_k +
	\begin{bmatrix}
		\mat{B} \cdot \mat{J}_{\theta_k}\\
		\mat{G}
	\end{bmatrix}
	U_k + 
	\begin{bmatrix}
		\rvec{w}_k \\
		\vec{0}_d
	\end{bmatrix} \ . \\ 
\end{align*}
Using (\ref{eq:LSchlange}) results in
\begin{align}
	\rvec{\psi}_{k+1} & = \underbrace{\left[
	\begin{bmatrix}
		\mat{A} & \mat{B} \cdot \mat{H}_{\theta_k}\\
		\mat{0}_{d\times s} & \mat{F}
	\end{bmatrix}
	- 
	\begin{bmatrix}
		\mat{B} \cdot \mat{J}_{\theta_k} \cdot \tilde{\mat{L}}\\
		\mat{G} \cdot \tilde{\mat{L}}
	\end{bmatrix}
	 \right]}_{\tilde{\mat{A}}_{\theta_k}} \rvec{\psi}_k + 
	\underbrace{\begin{bmatrix}
		\rvec{w}_k \\
		\vec{0}_d
	\end{bmatrix}}_{\tilde{\rvec{w}}_k}\\
	& = \tilde\mat{A}_{\theta_k} \rvec{\psi}_k + \tilde{\rvec{w}}_k  \ . \label{eq:closedLoopedSystem}
\end{align}
The closed-loop system described by (\ref{eq:closedLoopedSystem}) can be interpreted as an inhomogeneous Markovian jump linear system (MJLS). 
For this kind of system, several results on mean square stability are available in the literature, e.g.,  \cite{Costa05} and \cite{Feng92}. 
In the following, we adopt the concept of mean square stability and proof from \cite{Costa05}.
\begin{Definition}
	The system (\ref{eq:closedLoopedSystem}) with Markovian jump parameter $\theta_k$ is mean square stable (MSS), if for any initial condition $\theta_0 \in \left\{0, 1, \cdots, N\right\}$ and $\rvec{\psi}_0 \in \mathbb{R}^{d+s}$ there exist a bounded $\vec{\mu} \in \mathbb{R}^{d+s}$ and a symmetric positive-semidefinite matrix $\mat{M}$ (independent of $\rvec{\psi}_0$ and $\theta_0$) such that
	\begin{eqnarray}
		\lim\limits_{k\rightarrow \infty} E\left\{ \rvec{\psi}_k \right\} = \vec{\mu}\ , \\
		\lim\limits_{k\rightarrow \infty} E\left\{ \rvec{\psi}_k \rvec{\psi}_k^T\right\} = \mat{M} \ .
	\end{eqnarray}
\end{Definition}
\begin{Theorem} 
		The system (\ref{eq:closedLoopedSystem}) with Markovian jump parameter $\theta_k$ and transition matrix $\mat{P}$ is stable in the mean square sense, if and only if 
	\begin{eqnarray}
		r_\sigma((\mat{P}^T \otimes \mat{I}_{n^2})\cdot {\rm diag}[\tilde{\mat{A}}_i \otimes \tilde{\mat{A}}_i]) < 1 \ ,
	\end{eqnarray}
	where $r_\sigma\(\mat{M}\)$ is the spectral radius of $\mat{M}$ and ${\rm diag}\left[ \mat{S}_i\right]$ is the block diagonal matrix built by $\mat{S}_i$ in the diagonal with $i = \{0, 1, ..., N\}$ and zero everywhere else, i.e.,
	\begin{equation}
	{\rm diag}\left[\mat{S}_i\right] =
		\begin{bmatrix}
			\mat{S}_0 & \mat{0} & \cdots & \mat{0}\\
			\mat{0} & \mat{S}_1 & \cdots & \mat{0}\\
			\vdots & \vdots & \vdots & \vdots\\
			\mat{0} & \mat{0} & \cdots & \mat{S}_N
		\end{bmatrix} \ .
	\end{equation}
\end{Theorem}
\begin{Proof}
The result follows from theorem 3.9 and 3.33 in \cite{Costa05}.
\end{Proof}


\section{Simulation Results}
\noindent
In this section, we evaluate the presented method by means of simulations with an inverted pendulum on a cart, which is a classical benchmark for illustrating various control techniques.
A basic description of this experimental setup can be found, e.g., in \cite{Anderson89}.

\begin{table}[t]
\centering
 \begin{tabular}{lr}
 \toprule
  Mass of the cart & $0.5$ kg \\
 Mass of the pendulum & $0.5$ kg\\
 Friction of the cart & $0.1$ N/m/s \\
 Length to pendulum center of mass & $0.3$ m \\
  Inertia of the pendulum & $0.006 \: {\rm kg} \cdot \mathrm m^2$\\ 
 \bottomrule
\end{tabular}
\caption{Parameters of the inverted pendulum used in the simulations.}
\label{tab:parameters}
\end{table}

\subsection{Simulation Setup}
For realization of (\ref{eq:givenController}), we use a classical LQR controller \cite{Kwakernaak72}.
In Table~\ref{tab:parameters}, the simulation parameters of the inverted pendulum are shown. %
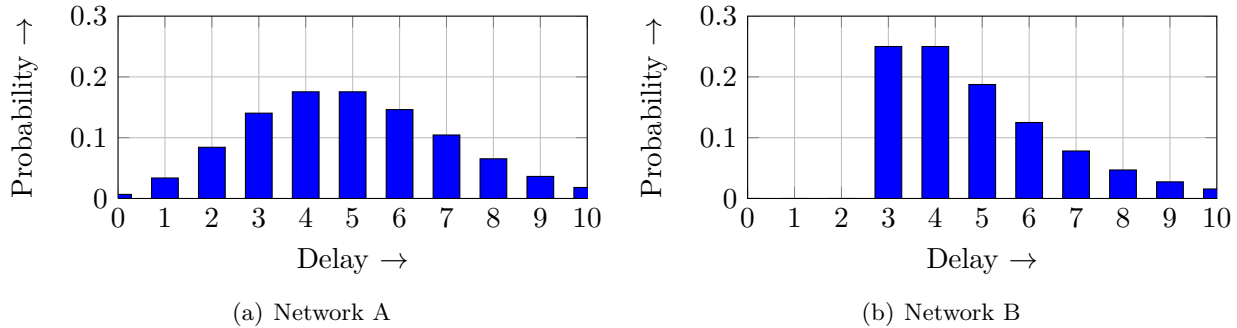
\begin{figure*}[ht!]
 \centering
 \subfigure[Network A]{
%
%
\begin{tikzpicture}

\begin{axis}[ymin=0, ymax=0.3, xmin=0, xmax=10, width=0.47\textwidth,
    height=4cm,
    xlabel=Delay $ \rightarrow$,
    ylabel=Probability $\rightarrow$, grid=major,
    xtick={0,1,2,3,4,5,6,7,8,9,10}]

\addplot[ybar,fill=blue,draw=black] plot coordinates { (0,0.00673795) (1,0.0336897) (2,0.0842243) (3,0.140374) (4,0.175467) (5,0.175467) (6,0.146223) (7,0.104445) (8,0.065278) (9,0.0362656) (10,0.0181328) };

\end{axis}
\end{tikzpicture}}
 \subfigure[Network B]{
%
%
\begin{tikzpicture}

\begin{axis}[ymin=0, ymax=0.3, xmin=0, xmax=10, width=0.47\textwidth,
    height=4cm,
    xlabel=Delay $ \rightarrow$,
    ylabel=Probability $\rightarrow$, grid=major,
    xtick={0,1,2,3,4,5,6,7,8,9,10}]

\addplot[ybar,fill=blue,draw=black] plot coordinates { (0,0) (1,0) (2,0) (3,0.25) (4,0.25) (5,0.1875) (6,0.125) (7,0.078125) (8,0.046875) (9,0.0273437) (10,0.015625) (11,0.00878906) (12,0.00488281) (13,0.00268555) (14,0.00146484) (15,0.000793457) (16,0.000427246) (17,0.000228882) (18,0.00012207) };

\end{axis}
\end{tikzpicture}}  
 \caption{Probability density functions over the time delays of the two networks considered in the evaluation.}
 \label{fig:densityNetwork}
\end{figure*}
The weighting matrices are chosen with
\begin{equation*}
  \mat{Q} = \left[ \begin{array}{cccc}
             5000 & 0 & 0 & 0 \\
	     0 & 0 & 0 & 0 \\
	     0 & 0 & 100 & 0 \\
	     0 & 0 & 0 & 0 \\
            \end{array} \right] \ ,
\end{equation*}
\begin{equation*}
 \mat{R} = 100 \ , 
\end{equation*}
and the continuous differential equation was sampled with a sampling time of 0.01~s. 
With this setting, the resulting state feedback matrix is 
\begin{equation*}
   \mat{L} =  \left[ -6.54,   -5.50,   28.72,   5.50 \right] \ .
\end{equation*}

At every time step $k$, we add a process noise $\rvec{w}_k$ to the position $\vec{x}_k$ of the cart and to the angle $\phi_k$ of the pendulum, which is characterized by a zero-mean Gaussian noise with varying standard deviation $\sigma_w$ for different simulation runs. 

For all simulation runs, the initial state vector $\vec{x}_0$ is
\begin{equation*}
	\vec{x}_0 = \left[ x_0, \dot{x}_0, \phi, \dot{\phi} \right]^T = \left[ 0,0.2, 0.2,0 \right]^T \ .
\end{equation*}

In order to simulate the transmission characteristics of the network, two probabilistic models for the occurring time delays were employed, whose probability density functions can be seen in Fig.~\ref{fig:densityNetwork}.
The time delays are bounded in both networks, because this allows a comparison to other sequence-based control methods, which need this assumption.

Overall, we conducted 100 Monte Carlo simulation runs for each combination of selected standard deviation of the process noise and selected probabilistic network model, where each run consists of 150 time steps.

\begin{figure*}[t]
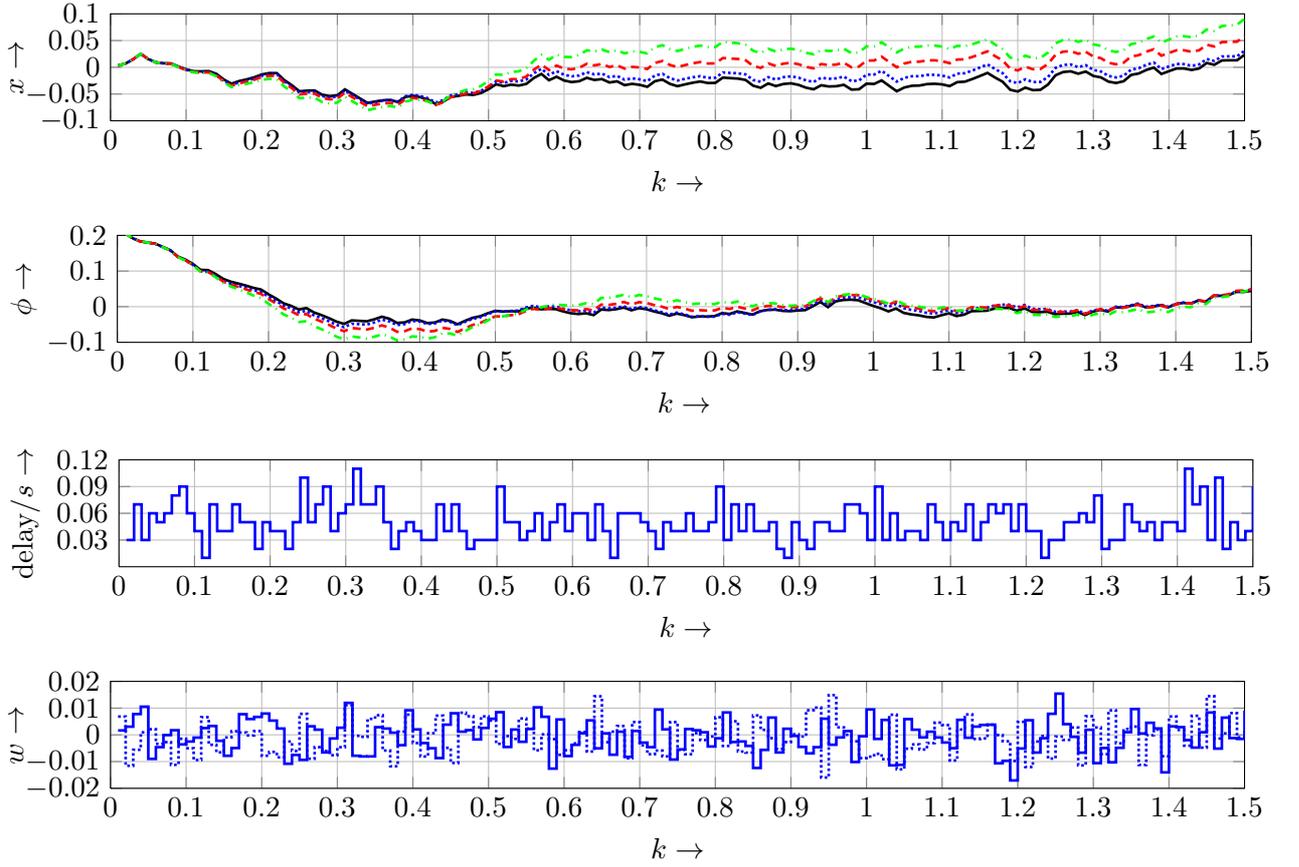

  \centering
  \subfigure{
    \input{figures/exampleuniformx.tikz}
  }
  \subfigure{
    \input{figures/exampleuniformphi.tikz}
  }
  \subfigure{
%
%
\begin{tikzpicture}

\definecolor{mycolor1}{rgb}{1,1,0}

\begin{axis}[ymin=0, ymax=0.12, xmin=0, xmax=1.5, width=\textwidth,
    height=3cm,
    xlabel=$k \rightarrow$,
    ylabel=$\text{delay}/s \rightarrow$, grid=major,
    ytick={0.03,0.06,0.09,0.12},
    scaled y ticks = false,
    y tick label style={/pgf/number format/fixed}]

\addplot[const plot,color=blue,solid, line width=1.0pt] plot coordinates { (0.01,0.03) (0.02,0.07) (0.03,0.03) (0.04,0.06) (0.05,0.05) (0.06,0.06) (0.07,0.08) (0.08,0.09) (0.09,0.06) (0.1,0.04) (0.11,0.01) (0.12,0.07) (0.13,0.04) (0.14,0.04) (0.15,0.07) (0.16,0.05) (0.17,0.05) (0.18,0.02) (0.19,0.05) (0.2,0.04) (0.21,0.04) (0.22,0.02) (0.23,0.05) (0.24,0.1) (0.25,0.05) (0.26,0.07) (0.27,0.09) (0.28,0.04) (0.29,0.06) (0.3,0.07) (0.31,0.11) (0.32,0.07) (0.33,0.07) (0.34,0.09) (0.35,0.05) (0.36,0.02) (0.37,0.04) (0.38,0.05) (0.39,0.04) (0.4,0.03) (0.41,0.03) (0.42,0.07) (0.43,0.04) (0.44,0.07) (0.45,0.04) (0.46,0.02) (0.47,0.03) (0.48,0.03) (0.49,0.03) (0.5,0.09) (0.51,0.05) (0.52,0.05) (0.53,0.03) (0.54,0.04) (0.55,0.06) (0.56,0.03) (0.57,0.05) (0.58,0.07) (0.59,0.04) (0.6,0.06) (0.61,0.06) (0.62,0.03) (0.63,0.07) (0.64,0.04) (0.65,0.01) (0.66,0.06) (0.67,0.06) (0.68,0.06) (0.69,0.05) (0.7,0.04) (0.71,0.02) (0.72,0.05) (0.73,0.04) (0.74,0.04) (0.75,0.06) (0.76,0.04) (0.77,0.04) (0.78,0.05) (0.79,0.09) (0.8,0.03) (0.81,0.07) (0.82,0.04) (0.83,0.07) (0.84,0.04) (0.85,0.04) (0.86,0.05) (0.87,0.02) (0.88,0.01) (0.89,0.05) (0.9,0.03) (0.91,0.02) (0.92,0.05) (0.93,0.05) (0.94,0.04) (0.95,0.04) (0.96,0.07) (0.97,0.07) (0.98,0.06) (0.99,0.03) (1,0.09) (1.01,0.03) (1.02,0.06) (1.03,0.03) (1.04,0.04) (1.05,0.04) (1.06,0.03) (1.07,0.07) (1.08,0.05) (1.09,0.03) (1.1,0.05) (1.11,0.04) (1.12,0.07) (1.13,0.03) (1.14,0.05) (1.15,0.07) (1.16,0.06) (1.17,0.07) (1.18,0.04) (1.19,0.07) (1.2,0.04) (1.21,0.04) (1.22,0.01) (1.23,0.03) (1.24,0.03) (1.25,0.05) (1.26,0.05) (1.27,0.06) (1.28,0.05) (1.29,0.08) (1.3,0.02) (1.31,0.03) (1.32,0.03) (1.33,0.07) (1.34,0.05) (1.35,0.04) (1.36,0.05) (1.37,0.03) (1.38,0.05) (1.39,0.05) (1.4,0.04) (1.41,0.11) (1.42,0.07) (1.43,0.09) (1.44,0.03) (1.45,0.1) (1.46,0.02) (1.47,0.05) (1.48,0.03) (1.49,0.04) (1.5,0.09) };

\end{axis}

\end{tikzpicture}
  }
  \subfigure{
%
%
\begin{tikzpicture}

\definecolor{mycolor1}{rgb}{1,1,0}

\begin{axis}[ymin=-0.02, ymax=0.02, xmin=0, xmax=1.5, width=\textwidth,
    height=3cm,
    xlabel=$k \rightarrow$,
    ylabel=$w \rightarrow$, grid=major,
    scaled y ticks = false,
    y tick label style={/pgf/number format/fixed}]

\addplot[const plot,color=blue,solid,line width=1.0pt] plot coordinates { (0.01,0.00164388) (0.02,0.00305499) (0.03,0.00777753) (0.04,0.0105428) (0.05,-0.00899645) (0.06,-0.00460819) (0.07,-9.86622e-005) (0.08,0.00172397) (0.09,-0.00357345) (0.1,-0.0025) (0.11,-0.00102873) (0.12,0.00253419) (0.13,0.00331038) (0.14,-0.00223786) (0.15,-0.00783757) (0.16,-0.00493263) (0.17,0.00771771) (0.18,0.00591179) (0.19,0.0066941) (0.2,0.00796721) (0.21,0.00532228) (0.22,0.00235746) (0.23,-0.0108192) (0.24,-0.00751168) (0.25,-0.00937158) (0.26,0.00319876) (0.27,0.00192167) (0.28,-0.00658161) (0.29,-0.00135166) (0.3,0.0026611) (0.31,0.011951) (0.32,-0.00782271) (0.33,-0.00808383) (0.34,-0.00806755) (0.35,0.00286203) (0.36,0.00235227) (0.37,0.000812496) (0.38,-0.00387503) (0.39,0.00918222) (0.4,0.00202888) (0.41,-0.00306987) (0.42,-0.00500474) (0.43,-0.00870422) (0.44,0.00380492) (0.45,0.00807377) (0.46,0.00119263) (0.47,0.00176033) (0.48,-0.00171815) (0.49,0.00257794) (0.5,0.00600688) (0.51,0.00846384) (0.52,-0.00454322) (0.53,0.00198175) (0.54,-0.00259926) (0.55,0.00219614) (0.56,0.0102775) (0.57,0.00675123) (0.58,-0.0126361) (0.59,0.00574649) (0.6,-0.00793707) (0.61,-0.00374147) (0.62,0.000606564) (0.63,0.00541056) (0.64,-0.00505104) (0.65,0.000562206) (0.66,-0.00743706) (0.67,0.0011664) (0.68,-0.00823423) (0.69,0.00145226) (0.7,-0.00290784) (0.71,-0.00461628) (0.72,0.00944817) (0.73,0.00204869) (0.74,-0.00924676) (0.75,-0.00543713) (0.76,0.00217778) (0.77,0.00170815) (0.78,0.000968954) (0.79,-0.0043641) (0.8,-0.0012912) (0.81,0.00836793) (0.82,-0.00268392) (0.83,-0.00248449) (0.84,-0.00415229) (0.85,-0.0123897) (0.86,-0.00562948) (0.87,0.00644747) (0.88,-0.0065962) (0.89,0.00227918) (0.9,-0.000388422) (0.91,-0.0045182) (0.92,-0.00692232) (0.93,-0.00460857) (0.94,0.00603522) (0.95,-0.00370757) (0.96,-0.000973437) (0.97,0.00142099) (0.98,-0.010148) (0.99,-0.000528881) (1,0.00949334) (1.01,-0.00152251) (1.02,0.00557743) (1.03,-0.0114141) (1.04,-0.0112239) (1.05,0.00588284) (1.06,0.00191958) (1.07,-0.000961174) (1.08,0.0011721) (1.09,-0.000974991) (1.1,-0.00286977) (1.11,-0.0050017) (1.12,0.00561678) (1.13,-0.00108508) (1.14,0.00307901) (1.15,0.00370386) (1.16,0.00380287) (1.17,-0.0106407) (1.18,-0.00971458) (1.19,-0.0170526) (1.2,-0.00510725) (1.21,0.00545664) (1.22,-0.00662475) (1.23,-0.00126271) (1.24,0.00787017) (1.25,0.0154097) (1.26,0.00399245) (1.27,-0.000753425) (1.28,-0.00666842) (1.29,0.00189321) (1.3,-0.0105913) (1.31,-0.00492627) (1.32,-0.010284) (1.33,-0.00485827) (1.34,0.000998753) (1.35,0.00965445) (1.36,0.000508976) (1.37,0.00451176) (1.38,0.00238573) (1.39,-0.0140377) (1.4,0.00317099) (1.41,0.00288268) (1.42,-0.0019635) (1.43,-0.00477147) (1.44,0.000967743) (1.45,0.00834474) (1.46,-0.00515542) (1.47,0.00643849) (1.48,-0.00121184) (1.49,-0.00152329) (1.5,0.00902573) };

\addplot[const plot,color=blue,densely dotted,line width=1.0pt] plot coordinates { (0.01,0.00683289) (0.02,-0.0115693) (0.03,-0.00558165) (0.04,0.000944526) (0.05,0.00209057) (0.06,-0.003358) (0.07,-0.00272791) (0.08,-0.0112243) (0.09,-0.00146638) (0.1,-0.00501982) (0.11,-0.00941435) (0.12,0.00659162) (0.13,-0.00371253) (0.14,-0.00654552) (0.15,-0.000973373) (0.16,0.00134085) (0.17,0.00087463) (0.18,-0.000241394) (0.19,0.00140743) (0.2,-0.00781641) (0.21,-0.00638189) (0.22,-0.00842434) (0.23,-0.00249762) (0.24,-0.0060406) (0.25,0.00785223) (0.26,-0.00892096) (0.27,-0.00459896) (0.28,-0.00605424) (0.29,-0.00761898) (0.3,-0.00461458) (0.31,0.0110691) (0.32,-0.000364265) (0.33,-0.000746642) (0.34,0.00597257) (0.35,0.00733695) (0.36,-0.0058705) (0.37,-0.0117562) (0.38,0.00122782) (0.39,0.00755492) (0.4,-0.00306965) (0.41,-0.00809312) (0.42,0.000627162) (0.43,0.00718742) (0.44,-0.00486234) (0.45,-0.0105218) (0.46,0.00827516) (0.47,0.00695196) (0.48,0.00286217) (0.49,0.00689105) (0.5,0.00533798) (0.51,-0.00334954) (0.52,-0.000470362) (0.53,-0.00151865) (0.54,0.00610495) (0.55,0.00235634) (0.56,-0.00578195) (0.57,0.000183744) (0.58,-0.0016099) (0.59,-0.00748315) (0.6,-0.00577751) (0.61,-0.00321665) (0.62,0.001056) (0.63,-0.00624024) (0.64,0.0144089) (0.65,-0.00275846) (0.66,0.00119687) (0.67,0.00330034) (0.68,-0.00879199) (0.69,0.00478987) (0.7,-0.0039502) (0.71,-0.00723646) (0.72,-0.00587257) (0.73,-0.00806023) (0.74,0.00477197) (0.75,-0.00562577) (0.76,-0.00725815) (0.77,0.000661429) (0.78,-0.00194143) (0.79,-0.000611545) (0.8,0.00656857) (0.81,-0.00415746) (0.82,0.0050792) (0.83,-0.00624163) (0.84,-0.00662767) (0.85,0.00727397) (0.86,0.00187592) (0.87,-0.00124644) (0.88,-0.00526371) (0.89,-0.00269651) (0.9,0.000200154) (0.91,-0.000627809) (0.92,0.00919996) (0.93,0.0101783) (0.94,-0.0159741) (0.95,0.0148589) (0.96,0.00279998) (0.97,2.18881e-005) (0.98,-0.00241778) (0.99,-0.00870703) (1,-0.0081637) (1.01,-0.00710187) (1.02,-0.00909951) (1.03,0.00752701) (1.04,-0.0128976) (1.05,-0.00517863) (1.06,-0.00357073) (1.07,-0.0054907) (1.08,-0.00308146) (1.09,0.00457564) (1.1,0.0036518) (1.11,-0.0095974) (1.12,0.00199392) (1.13,0.00762701) (1.14,-0.00153995) (1.15,-0.00134945) (1.16,0.00378921) (1.17,-8.58125e-005) (1.18,-0.00239578) (1.19,-0.00977271) (1.2,0.00487337) (1.21,-0.0120313) (1.22,-0.00249638) (1.23,-0.000426008) (1.24,-0.00888132) (1.25,-0.000654639) (1.26,-0.00188595) (1.27,0.00273961) (1.28,-0.011428) (1.29,0.00696393) (1.3,0.00275299) (1.31,-0.00721506) (1.32,0.0091622) (1.33,-0.0031692) (1.34,0.00218419) (1.35,0.00308153) (1.36,-0.0112404) (1.37,-0.00013447) (1.38,0.00195478) (1.39,-0.00799617) (1.4,0.00686556) (1.41,0.00182202) (1.42,-0.00405928) (1.43,0.00712555) (1.44,-0.00478177) (1.45,0.0144665) (1.46,-0.00138593) (1.47,0.000541159) (1.48,0.00813324) (1.49,-0.000935926) (1.5,0.00208261) };

\end{axis}

\end{tikzpicture}
  }
  \caption{Example state trajectory for a single simulation run with network A and a standard deviation $\sigma_w = 0.006$ of the process noise. 
    The identifier $x$ denotes the position of the cart and $\phi$ the angle of the pendulum. 
    The result of the controller without a network (CS) is depicted with a solid black line (\ref{pgfplots:CS}), the proposed approach VCI-NCS with a dotted blue (\ref{pgfplots:VI-NCS}),
   OL-NCS with a dashed red (\ref{pgfplots:OL-NCS}), and PC-NCS with a dashed dotted green line (\ref{pgfplots:PC-NCS}).}
  \label{fig:exampleTrajectory}
\end{figure*}

\begin{table}[t]
\centering
 \begin{tabular}{l||cccc}
 \toprule
 & CS & VCI-NCS & OL-NCS & PC-NCS \\
 \hline
 $\sigma_w = 0.001$ / network A & 3.20  & 3.90 & 3.78 & 5.18 \\
 $\sigma_w = 0.003$ / network A & 12.78 & 17.15 & 18.51 & 36.44 \\
 $\sigma_w = 0.006$ / network A & 45.93 & 62.74 & 67.24 & 144.15 \\
 $\sigma_w = 0.009$ / network A & 98.57 & 137.38 & 140.02 & 262.28 \\
 $\sigma_w = 0.012$ / network A & 186.60 & 275.22 & 290.18 & 619.88 \\ 
 $\sigma_w = 0.001$ / network B & 3.03 & 3.42 & 3.34 & 5.37 \\
 $\sigma_w = 0.003$ / network B & 88.85 & 129.68 & 141.26 & 317.16 \\
 $\sigma_w = 0.006$ / network B & 110.37 & 164.32 & 166.49 & 359.85 \\
 $\sigma_w = 0.009$ / network B & 122.50 & 158.54 & 175.45 & 386.63 \\
 $\sigma_w = 0.012$ / network B & 197.94 & 275.24 & 310.50 & 588.27 \\ 
\bottomrule
\end{tabular}
\caption{Cumulated costs of the simulated state trajectories averaged over the 100 Monte Carlo runs for non-networked controller (CS), the proposed method considering virtual control inputs (VCI-NCS), the NCS method sending open loop control inputs (OL-NCS), and the approach based on prediction consistency (PC-NCS).}
\label{tab:results}
\end{table}

We compare the presented technique for NCS with virtual control inputs (VCI-NCS) to three other NCS control approaches.
For better analyzing the quality of the compensation technique for time delays, we consider a classical LQR (abbreviated by CS) with a transparent connection between controller and actuator.
In this case, all calculated control inputs 
\begin{equation*}
 \vec{u}_{k \vert k } = \mat{L} \cdot \vec{x}_{k \vert k}
\end{equation*}
are received  by the actuator without any time delay. 
The control quality of CS can be seen as a ground truth for the NCS control methods.

Furthermore, we compare VCI-NCS to a widely used NCS controller that sends at every time step a sequence of control inputs resulting from an open-loop control problem (OL-NCS) 
\cite{Quevedo11, Quevedo07}.
In more detail, the packet $U_k$ sent in time step $k$ contains the entries 
\begin{eqnarray*}
 \vec{u}_{k \vert k} &=& \mat{L} \cdot \vec{x}_{k \vert k} \ ,\\
 \vec{u}_{k+1 \vert k} &=& \mat{L} \cdot \vec{x}_{k+1 \vert k} \ ,\\
 &\vdots&\\
 \vec{u}_{k+N-1 \vert k} &=& \mat{L} \cdot \vec{x}_{k+N-x \vert k} \ , \\ 
\end{eqnarray*}
where $\vec{x}_{k+i \vert k}$ is determined for $1 \leq i < N$ according to
\begin{equation*}
 \vec{x}_{k+i\vert k} = \mat{A} \vec{x}_{k+i-1\vert k} + \mat{B} \vec{u}_{ k+i-1 \vert k} \ .
\end{equation*}

Finally, we implemented an instance of a class of NCS approaches that ensures the so-called prediction consistency by means of a deterministic protocol between actuator and controller.
Since we assume that successfully transmitted packets are not acknowledged by the network protocol, the parameter $\tau_{\text{max}}$ of the PC-NCS approach described in \cite{Gruene09} was set to the true maximal delay.

\subsection{Results}
In Fig.~\ref{fig:exampleTrajectory}, an example state trajectory of a simulation run with network A and a standard deviation $\sigma_w = 0.006$ is depicted.
The state trajectory of VCI-NCS is very similar to the one generated by the non-networked controller. 
Thus, the proposed method is an adequate technique for compensating time delays.
In contrast, the trajectories of OL-NCS and PC-NCS are outperformed by VCI-NCS.

In order to make quantitative statements, we conducted 100 Monte Carlo simulation runs with different parameter settings. 
The averaged costs over all runs are shown in Table~\ref{tab:results}.
For small system noise, the cumulated averaged costs of the proposed approach
using virtual inputs and the NCS method sending open loop sequences are very similar.
This is based on the fact that in this case, the components of the Dirac Mixture densities characterizing the virtual control inputs do not differ strongly, or more precisely,
the variances of these densities are very small. 
Thus, both methods use control inputs for generating the input sequences, which are very similar to the actually applied input.
In contrast, if the system noise increases, the variance of the Dirac Mixture densities also increases.
Considering the NCS approach sending open loop sequences, inputs used for the prediction of the system evolution do not coincide in general with the actually applied inputs.
As a result, the quality of control decreases. 
By incorporating the potentially applied inputs in a stochastic way, the network-induced disturbances can be compensated better, even if the system noise is large.


\section{Conclusions} \label{Sec_Conclusions}
\noindent
We presented a novel sequence-based predictive control scheme for NCS that extends a given feedback controller to explicitly incorporate communication aspects as transmission delays and packet losses.
The key idea of our approach is that the controller subsumes its knowledge about the control inputs potentially applied by the actuator in  form of a discrete probability density function, 
the so-called \emph{virtual control inputs}.
Based on this probabilistic description, the controller determines sequences of future control inputs which are sent to the actuator.

To the best of our knowledge, the concept of virtual control inputs is innovative and promising, 
especially since simulation results with an inverted pendulum show an excellent performance of the proposed approach in comparison to standard NCS methods.

Future work will be concerned with incorporation of  further information into the control decision. For example, promising aspects might be the investigation of
\begin{enumerate}
 \item allowing for time-varying transmission characteristics of the network resulting in more adequate models of the real system,
 \item acknowledgements of successfully transmitted packets which allow to reduce the components of the virtual control inputs,
 \item closed-loop stability if the weighting factors of the virtual control inputs are estimated by the Wonham filter.
\end{enumerate}

\section*{Appendix: Derivation of the Transition Matrix}
The entries $p_{(i, \ j)}$ (with $i, j \in \{0, 1, \cdots, N + 1$) of the transition matrix can be categorized into three groups. The first group consists of the entries with $j \geq i + 2$, i. e., entries that are in the upper right triangle of $\mat{P}$, which describe transitions from $\theta_k = i$ to $\theta_{k+1} = j$. These entries have to be zero as ${\theta}_k$, i.e. the age of the buffered control sequence, can only increase by one per time step due to the buffering scheme.

The second group consists of the entries in the upper diagonal, i.e., the entries $p_{(i, \ i + 1)}$, that describe the probability that the age of the buffered control sequence will increase by one. This corresponds to the case that the buffered sequence at time step $k$ is not replaced and stays in the buffer. This will only occur if the actuator does not receive a packet that was generated after the actually buffered sequence. It therefore holds that
\begin{eqnarray*}
	p_{(i,\  i+1)} &=& \prod\limits_{j=0}^{i} (1-\check{q}_{j}) \ ,
	\label{eq:pii}
\end{eqnarray*}
where $\check{q}_{j}$ denotes the probability that a packet that was generated $j$ time steps ago and \emph{that has not been received yet}, will be received during the next time step. It holds that
\begin{equation}
	\check{q}_{j} = {q_j} \left(\displaystyle\sum_{r=j}^{N + 1}q_r \ \right)^{-1}
	\label{eq:qi}
\end{equation}
where $q_i$ describes the probability of the event that a packet is delayed by $i$ time steps, which can easily be calculated with the probability density function of the time delays given. Therefore, the first term in Eq.(\ref{eq:qi}) describes the probability that the sequence generated $i$ time steps ago will arrive during the next time step ignoring that we know that the sequence has not been received yet. The second term normalizes this probability by the probability that this sequence will arrive at all. Combining Eq.(\ref{eq:qi}) and Eq.(\ref{eq:pii}) it holds
\begin{eqnarray}
	p_{(i,\  i+1)} &=& \prod\limits_{j=0}^{i} (1-\check{q}_{j}) 
	= \prod\limits_{j=0}^{i} \left(1 - {q_j} \left(\displaystyle\sum_{r=j}^{N + 1}q_r \ \right)^{-1}\right) 
	= \prod\limits_{j=0}^{i} \left(
	\frac
	{ 
		\displaystyle\sum_{r=j+1}^{N + 1} q_r 
	} 
	{ 
		\displaystyle\sum_{r=j}^{N + 1} q_r 
	} \right) \nonumber \\
	&=& \frac{\displaystyle\sum_{r=i+1}^{N + 1} q_r } { \displaystyle\sum_{r=0}^{N + 1} q_r } =  \displaystyle\sum_{r=i+1}^{N + 1} q_r = 1 - \displaystyle\sum_{r=0}^{i} q_r \  .
	\label{eq:pii2}
\end{eqnarray}

The remaining entries $p_{(i, \ j)}$ with $j\leq i$ build the lower triangle of $\mat{P}$ and describe transitions where $\theta_k$ does not increase. 
This corresponds to the case when the buffered sequence is replaced by a newer one.
The probability that this happens is given by the probability that a new packet is received (that was generated after the buffered sequence) and all packets generated after this new packet are not received. This means
\begin{eqnarray}
	p_{(i,\  j)} &=&  \check{q}_j \prod\limits_{r=0}^{j-1} (1-\check{q}_{r})
\end{eqnarray}
Note that in the above equation the probabilities whether a packet is received in the next time step is conditioned on the information that the corresponding packet has not been received yet. This condition is necessary since, if neglected, the transition could not start in state $\theta_k = i$. Using Eq. (\ref{eq:qi}) and Eq. (\ref{eq:pii2}) it holds that
\begin{eqnarray}
	p_{(i,\  j)} &=&  \check{q}_j \prod\limits_{r=0}^{j-1} (1-\check{q}_{r}) = {q_j} \left(\displaystyle\sum_{r=j}^{N + 1}q_r \ \right)^{-1} \displaystyle\sum_{r=j}^{N + 1} q_r = q_j \ .
\end{eqnarray}

    

\bibliographystyle{IEEEtran}
\bibliography{literature}

\begin{thebibliography}{10}
\providecommand{\url}[1]{#1}
\csname url@samestyle\endcsname
\providecommand{\newblock}{\relax}
\providecommand{\bibinfo}[2]{#2}
\providecommand{\BIBentrySTDinterwordspacing}{\spaceskip=0pt\relax}
\providecommand{\BIBentryALTinterwordstretchfactor}{4}
\providecommand{\BIBentryALTinterwordspacing}{\spaceskip=\fontdimen2\font plus
\BIBentryALTinterwordstretchfactor\fontdimen3\font minus
  \fontdimen4\font\relax}
\providecommand{\BIBforeignlanguage}[2]{{%
\expandafter\ifx\csname l@#1\endcsname\relax
\typeout{** WARNING: IEEEtran.bst: No hyphenation pattern has been}%
\typeout{** loaded for the language `#1'. Using the pattern for}%
\typeout{** the default language instead.}%
\else
\language=\csname l@#1\endcsname
\fi
#2}}
\providecommand{\BIBdecl}{\relax}
\BIBdecl

\bibitem{Yang06}
T.~Yang, ``Networked {C}ontrol {S}ystem: a {B}rief {S}urvey,'' \emph{IEEE
  Proceedings, Control Theory and Applications}, vol. 153, no.~4, pp. 403--412,
  July 2006.

\bibitem{Hespanha07}
J.~Hespanha, P.~Naghshtabrizi, and Y.~Xu, ``A {S}urvey of {R}ecent {R}esults in
  {N}etworked {C}ontrol {S}ystems,'' \emph{Proceedings of the IEEE}, vol.~95,
  no.~1, pp. 138--162, 2007.

\bibitem{Seiler05}
P.~Seiler and R.~Sengupta, ``An {H} $\infty$ {A}pproach to {N}etworked
  {C}ontrol,'' \emph{IEEE Transactions on Automatic Control}, vol.~50, no.~3,
  pp. 356--364, 2005.

\bibitem{Hirche04}
S.~Hirche and M.~Buss, ``Packet {L}oss {E}ffects in {P}assive {T}elepresence
  {S}ystems,'' in \emph{Proceedings of the 43rd IEEE Conference on Decision and
  Control}, vol.~4.\hskip 1em plus 0.5em minus 0.4em\relax IEEE, 2004, pp.
  4010--4015.

\bibitem{Ogren04}
P.~Ogren, E.~Fiorelli, and N.~Leonard, ``Cooperative {C}ontrol of {M}obile
  {S}ensor {N}etworks: {A}daptive {G}radient {C}limbing in a {D}istributed
  {E}nvironment,'' \emph{IEEE Transactions on Automatic Control}, vol.~49,
  no.~8, pp. 1292--1302, 2004.

\bibitem{Zhang01}
W.~Zhang, M.~Branicky, and S.~Phillips, ``Stability of {N}etworked {C}ontrol
  {S}ystems,'' \emph{IEEE Control Systems Magazine}, vol.~21, no.~1, pp.
  84--99, 2001.

\bibitem{Bemporad10}
A.~Bemporad, M.~Heemels, and M.~Johansson, \emph{Networked {C}ontrol
  {S}ystems}.\hskip 1em plus 0.5em minus 0.4em\relax Springer-Verlag New York
  Inc, 2010, vol. 406.

\bibitem{Heemels10}
W.~Heemels, A.~Teel, N.~van~de Wouw, and D.~Nesic, ``Networked {C}ontrol
  {S}ystems {W}ith {C}ommunication {C}onstraints: Tradeoffs {B}etween
  {T}ransmission {I}ntervals, {D}elays and {P}erformance,'' \emph{IEEE
  Transactions on Automatic Control}, vol.~55, no.~8, pp. 1781--1796, 2010.

\bibitem{Gruene09}
L.~Gr\"une, J.~Pannek, and K.~Worthmann, ``A {P}rediction {B}ased {C}ontrol
  {S}cheme for {N}etworked {S}ystems with {D}elays and {P}acket {D}ropouts,''
  in \emph{Proceedings of the 48th IEEE Conference on Decision and Control held
  jointly with the 28th Chinese Control Conference}.\hskip 1em plus 0.5em minus
  0.4em\relax IEEE, 2009, pp. 537--542.

\bibitem{Polushin08}
I.~Polushin, P.~Liu, and C.~Lung, ``On the {M}odel-{B}ased {A}pproach to
  {N}onlinear {N}etworked {C}ontrol {S}ystems,'' \emph{Automatica}, vol.~44,
  no.~9, pp. 2409--2414, 2008.

\bibitem{Quevedo08}
D.~Quevedo, E.~Silva, and G.~Goodwin, ``Control over {U}nreliable {N}etworks
  {A}ffected by {P}acket {E}rasures and {V}ariable {T}ransmission {D}elays,''
  \emph{IEEE Journal on Selected Areas in Communications}, vol.~26, no.~4, pp.
  672--685, 2008.

\bibitem{Quevedo11}
D.~Quevedo and D.~Nesic, ``Input-to-{S}tate {S}tability of {P}acketized
  {P}redictive {C}ontrol over {U}nreliable {N}etworks {A}ffected by
  {P}acket-{D}ropouts,'' \emph{IEEE Transactions on Automatic Control}, no.~99,
  pp. 1--1, 2011.

\bibitem{Quevedo07}
D.~Quevedo, E.~Silva, and G.~Goodwin, ``Packetized {P}redictive {C}ontrol over
  {E}rasure {C}hannels,'' in \emph{Proceedings of the American Control
  Conference 2009}.\hskip 1em plus 0.5em minus 0.4em\relax IEEE, 2007, pp.
  1003--1008.

\bibitem{Liu07}
G.~Liu, Y.~Xia, J.~Chen, D.~Rees, and W.~Hu, ``Networked {P}redictive {C}ontrol
  of {S}ystems with {R}andom {N}etwork {D}elays in both {F}orward and
  {F}eedback {C}hannels,'' \emph{IEEE Transactions on Industrial Electronics},
  vol.~54, no.~3, pp. 1282--1297, 2007.

\bibitem{Bemporad98}
A.~Bemporad, ``Predictive {C}ontrol of {T}eleoperated {C}onstrained {S}ystems
  with {U}nbounded {C}ommunication {D}elays,'' in \emph{Proceedings of the 37th
  IEEE Conference on Decision and Control 1998}, vol.~2.\hskip 1em plus 0.5em
  minus 0.4em\relax IEEE, 1998, pp. 2133--2138.

\bibitem{Liu06}
G.~Liu, J.~Mu, D.~Rees, and S.~Chai, ``Design and {S}tability {A}nalysis of
  {N}etworked {C}ontrol {S}ystems with {R}andom {C}ommunication {T}ime {D}elay
  {U}sing the {M}odified mpc,'' \emph{International Journal of Control},
  vol.~79, no.~4, pp. 288--297, 2006.

\bibitem{Tang07}
P.~Tang and C.~De~Silva, ``Stability {V}alidation of a {C}onstrained {M}odel
  {P}redictive {N}etworked {C}ontrol {S}ystem with {F}uture {I}nput
  {B}uffering,'' \emph{International Journal of Control}, vol.~80, no.~12, pp.
  1954--1970, 2007.

\bibitem{Wonham64}
W.~Wonham, ``Some {A}pplications of {S}tochastic {D}ifferential {E}quations to
  {O}ptimal {N}onlinear {F}iltering,'' \emph{SIAM Journal on Control}, vol.~2,
  no.~3, pp. 347--369.

\bibitem{Costa05}
O.~do~Valle~Costa, M.~Fragoso, and R.~Marques, \emph{Discrete-{T}ime {M}arkov
  {J}ump {L}inear {S}ystems}.\hskip 1em plus 0.5em minus 0.4em\relax Springer
  Verlag, 2005.

\bibitem{Feng92}
X.~Feng, K.~Loparo, Y.~Ji, and H.~Chizeck, ``Stochastic {S}tability
  {P}roperties of {J}ump {L}inear {S}ystems,'' \emph{IEEE Transactions on
  Automatic Control}, vol.~37, no.~1, pp. 38--53, 1992.

\bibitem{Anderson89}
C.~Anderson, ``Learning to {C}ontrol an {I}nverted {P}endulum using {N}eural
  {N}etworks,'' \emph{IEEE Control Systems Magazine}, vol.~9, no.~3, pp. 31
  --37, apr 1989.

\bibitem{Kwakernaak72}
H.~Kwakernaak and R.~Sivan, \emph{Linear {O}ptimal {C}ontrol {S}ystems}.\hskip
  1em plus 0.5em minus 0.4em\relax Wiley-Interscience New York, 1972, vol. 172.

\end{thebibliography}

\end{document}